\documentclass[graybox]{svmult}
\usepackage{mathptmx}       
\usepackage{helvet}         
\usepackage{courier}        
\usepackage{makeidx}         
\usepackage{graphicx}        
\usepackage{multicol}        
\usepackage[bottom]{footmisc}
\usepackage{amsmath,amssymb,amsfonts}        
\usepackage{enumitem}
\makeindex             

\begin{document}
\title*{Thermodynamically Stable Knots in Semiflexible Polymers}
\author{Wolfhard Janke, Suman Majumder, Martin Marenz, Subhajit Paul} 
\institute{
Wolfhard Janke 
\at Institut f\"ur Theoretische Physik, Universit\"at Leipzig, IPF 231101, 
04081 Leipzig, Germany, \email{wolfhard.janke@itp.uni-leipzig.de} \and
Suman Majumder 
\at Amity Institute of Applied Sciences, Amity University Uttar Pradesh, 
Noida 201313, India, \email{suman.jdv@gmail.com} \and
Martin Marenz 
\at Institut f\"ur Theoretische Physik, Universit\"at Leipzig, IPF 231101, 
04081 Leipzig, Germany, \email{martin.marenz@itp.uni-leipzig.de} \and
Subhajit Paul 
\at International Center for Theoretical Sciences, Tata Institute of Fundamental 
Research, Bangalore 560089, India, \email{subhajit.paul@icts.res.in}
}

%
%
\maketitle

\abstract{Semiflexible polymers are widely used as a
paradigm for understanding structural phases in biomolecules
including folding of proteins. 
Here, we compare bead-spring and bead-stick variants of 
coarse-grained semiflexible polymer models
that cover the whole range from flexible to stiff
by conducting extensive replica-exchange Monte Carlo 
computer simulations. In the data analysis we focus on knotted 
conformations whose stability is shown to depend on the ratio 
$r_b/r_{\rm min}$ with $r_b$ denoting the equilibrium bond
length and $r_{\rm min}$ the distance of the strongest nonbonded
interactions. For both models, 
our results provide evidence that 
at low temperatures
for $r_b/r_{\rm min}$ outside
a small range around unity one always encounters 
knots as generic stable phases along with the usual frozen and 
bent-like structures.
%
%
By varying the bending stiffness, we observe rather strong
first-order-like structural transitions between the coexisting
phases characterized by these geometrically different motifs. 
Through analyses of 
the energy distributions close to the transition 
point,
we present exploratory estimates of the free-energy barriers 
between the coexisting phases.
%
%
%
%
%
} 

\section{Introduction}
\label{sec:intro}

\begin{figure}[b]
\begin{center}
\includegraphics[scale=.60]{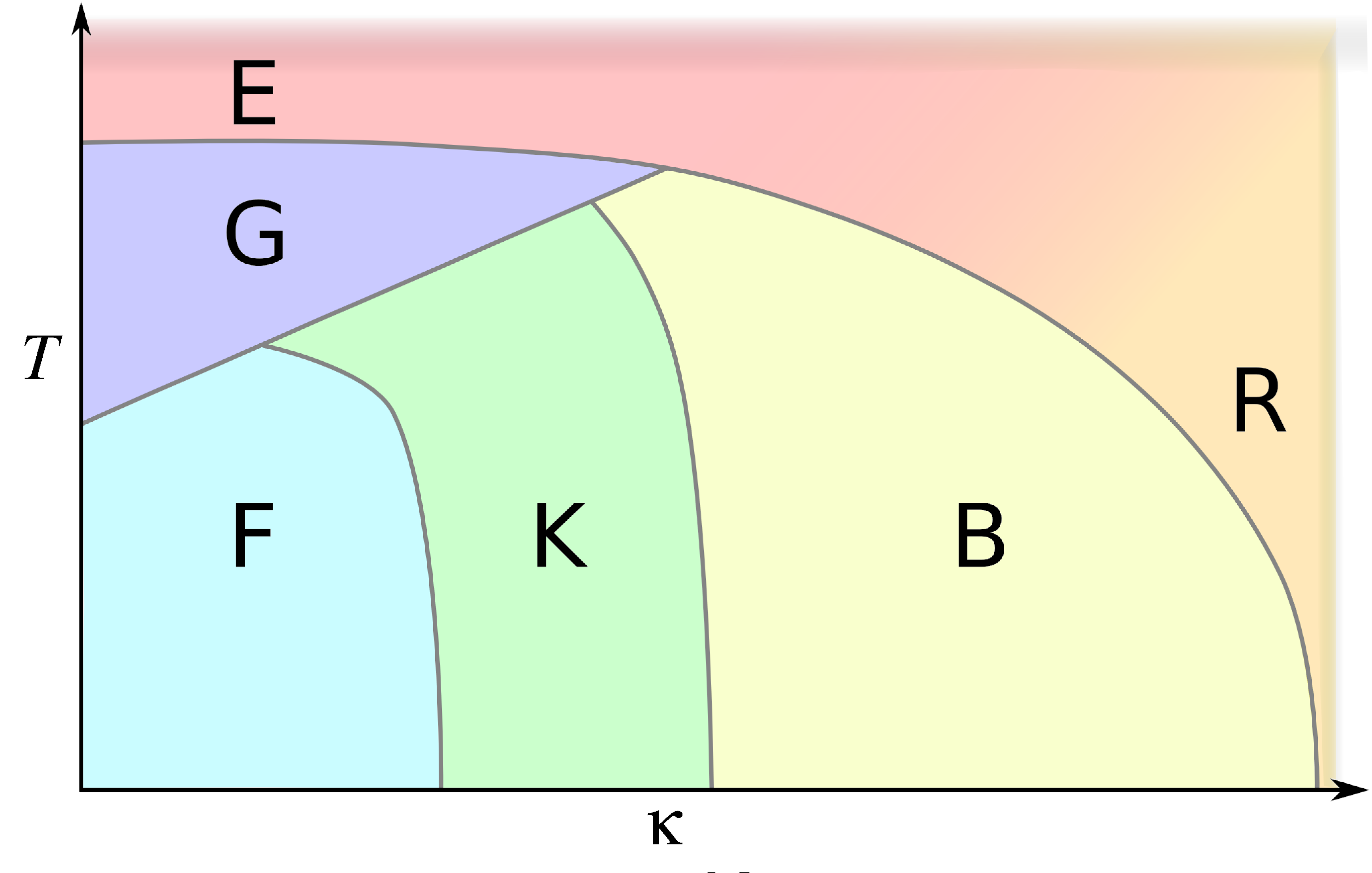}
\caption{
Generic phase diagram for a semiflexible polymer in the ($T,\kappa$)
plane. The labels stand short for
E: Extended coils,
R: Rods
G: Globules,
F: Frozen,
K: Stable knot phase,
B: Bent.
}
\label{fig:generic_phase-diagram}
\end{center}
\end{figure}

Computer simulation studies of atomistic macromolecular systems are often
limited by the accessible time scales. One therefore often considers less 
expensive coarse-grained models that are sufficient for understanding the 
generic features of macromolecules \cite{muller2002coarse}. This approach
can be 
classified
into bead-spring and bead-stick polymer models. In addition
to excluded volume effects and interactions among 
the monomers or beads, 
for semiflexible biopolymers one also has to take into account bending 
stiffness of the chain \cite{kratky1949}. With this motivation in mind,
Seaton \textit{et al.}\ \cite{seaton2013flexible} explored different phases 
(coiled, collapsed, frozen, bent, hairpin, and toroidal conformations) of a 
bead-spring 
model by tuning the bending stiffness $\kappa$. Similar phases show up in a 
bead-stick
model \cite{marenz2016knots} where in addition one 
observes 
at low temperatures $T$
also phases 
dominated
by thermodynamically stable knotted structures. 
For a sketch of the generic 
($T,\kappa$)
phase diagram, see Fig.\ \ref{fig:generic_phase-diagram}.
In Ref.\ \cite{marenz2016knots} it has been conjectured that the propensity for 
forming knots may depend on the ratio of the equilibrium bond length $r_b$ to 
the distance $r_{\rm{min}}$ where the nonbonded interaction potential is at its
minimum, i.e., the attractive interaction is strongest.

To shed more light onto the puzzle why knots apparently appear only in one of
the two models, we recently performed extensive comparative Monte Carlo 
simulations \cite{macromolecules21}. Our numerical results provide evidence that for 
both model variants, in certain ranges of bending stiffness, knotted 
structures are always observed when $r_b/r_{\rm{min}}$ is away from a small 
region around unity. Analytically, this can be understood by the competition 
of the nonbonded interaction and bending energy when minimizing the total energy.

\section{Models and Simulation}
\label{sec:models}

In both model 
variants, spherical beads with diameter $\sigma$
are considered as the monomers and the nonbonded interaction energy 
among them is taken 
for a polymer chain of length $N$
as
%
\begin{eqnarray}\label{potential_OLM}
E_{\rm nb}(r_{ij})=\sum_{i=1}^{N-2}\sum_{j=i+2}^{N} \left[ E_{\rm {LJ}}({\rm{min}}\{r_{ij},r_c\})-E_{\rm {LJ}}(r_c) \right]
\label{eq:E_nb}
\end{eqnarray}
with the Lennard-Jones (LJ) potential
\begin{eqnarray}\label{our_LJ}
E_{\rm {LJ}}(r_{ij})=4\epsilon \left[ \left( \frac{\sigma}{r_{ij}} \right)^{12} - \left (\frac{ \sigma}{r_{ij}} \right )^{6} \right]
\label{eq:E_LJ}
\end{eqnarray}
where $r_{ij}$
denotes the distance between the beads. The parameters
$\sigma$ and $\epsilon = 1$ set the length and energy scales,
respectively. Both 
$E_{\rm nb}$ and $E_{\rm LJ}$ have a minimum at $r_{\rm min} = 2^{1/6} \sigma$ 
with $E_{\rm LJ, min} = -\epsilon$. To be consistent with 
Ref.\ \cite{seaton2013flexible}, for the bead-spring model we employ
a cutoff distance $r_c=2.5\sigma$ and set $\sigma=2^{-1/6}$, implying 
$r_{\rm min} = 1$, whereas following our earlier work \cite{marenz2016knots}, 
for the bead-stick
model we do not apply any cutoff distance and set $\sigma=1$ such that
$r_{\rm min} = 2^{1/6}$. 
To model the 
bonds by springs
between successive beads in the bead-spring model,
we use the standard finitely extensible non-linear elastic 
(FENE) potential \cite{milchev1993off,milchev2001formation}
\begin{eqnarray}\label{FENE}
E_{\rm FENE}=-\frac{K}{2}R^2 \sum_{i=1}^{N-1}\ln\left[1-\left(\frac{r_{ii+1}-r_b}{R}\right)^2\right]
\end{eqnarray}
with $r_b$ denoting the equilibrium bond length where $E_{\rm{FENE}}$ is minimal,
$R=0.3$, and $K=40$.
The rigid bonds or sticks
in the bead-stick model 
are assumed to have a fixed bond length $r_b$, i.e., here
formally $E_{\rm FENE} \equiv 0$. The sum of nonbonded and bonded interactions 
(apart from bending stiffness) is denoted by $E_0 = E_{\rm nb} + E_{\rm FENE}$

The bending stiffness is introduced via the well-known
discretized worm-like chain cosine potential
\begin{equation}\label{stiff}
 E_{\rm{bend}} \equiv \kappa E_1 = \kappa \sum_{i=1}^{N-2}(1-\cos \theta_i)
\end{equation}
where $\theta_i$ is the angle between consecutive bonds and the model parameter
$\kappa$ controls the effective bending stiffness of the polymer.
The total energy of the semiflexible polymer is then
$E = E_0 + E_{\rm bend} = E_0 + \kappa E_1$.

While this model still looks rather simple, its simulation demands application 
of relatively complex Monte Carlo (MC) methods \cite{janke2018macromolecule}. 
Previously \cite{marenz2016knots} we have used a parallelized version of the 
multicanonical algorithm 
\cite{berg1991multicanonical,zierenberg2013scaling,janke2016SM} 
along with replica exchange (RE) (also known as parallel tempering) \cite{hukushima1996exchange}
and validated the results with the two-dimensional replica exchange method (2D-RE). 
In Ref.\ \cite{macromolecules21} we restricted ourselves to the 2D-RE algorithm. The set
of attempted MC updates included the usual crank-shaft, spherical-rotation, and pivot moves 
\cite{Austin2018}.

\begin{figure}[tb]
\begin{center}
\includegraphics[width=1.0\textwidth]{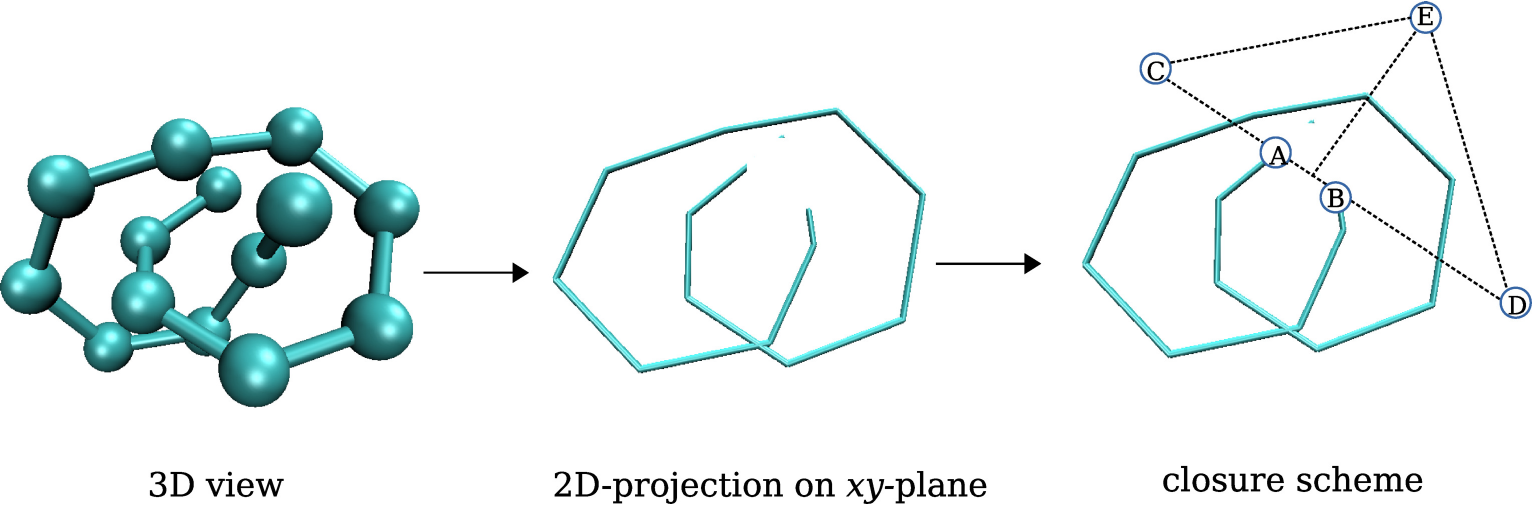}
\caption{
Closure scheme for an open polymer conformation in three-dimensional space
(3D view) applied to its two-dimensional projection onto one plane of a
Cartesian coordinate system (2D projection on $xy$-plane).
Taken from Ref.\ \cite{macromolecules21}.
\label{fig:closure_scheme}
}
\end{center}
\end{figure}

The definition of knots in an open polymer chain requires some care. 
Mathematically, knots can be defined only for closed curves \cite{kauffman1991}.
One therefore has first
to close an open polymer chain virtually for which several
rules are possible. We employed the specific closure prescription as discussed
in Refs.\ \cite{virnau2005knots,virnau2010,marenz2016knots,janke2016stable}
and depicted in Fig.\ \ref{fig:closure_scheme}.

\begin{figure}[tbh]
\begin{center}
\includegraphics[scale=.41]{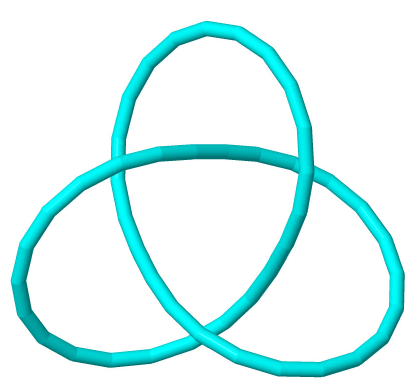}\hspace*{3.7mm}\
\includegraphics[scale=.47]{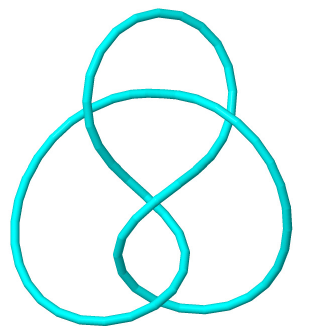}\hspace*{2.7mm}\
\includegraphics[scale=.36]{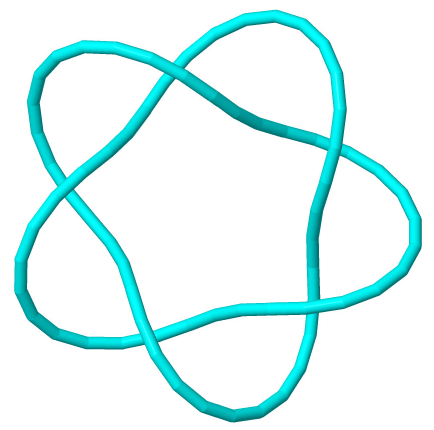}\hspace*{2.7mm}\
\includegraphics[scale=.38]{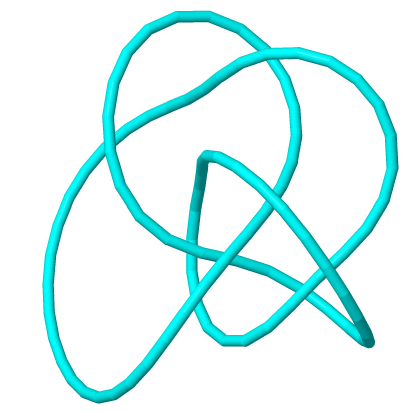}
\caption{
Sketch of the knot type 3$_1$ characterizing the stable knot phase K3$_1$
for semiflexible polymers of length $N=14$ and the knot types
4$_1$, 5$_1$, and 8$_{19}$ occurring for $N=28$.
}
\label{fig:knots}
\end{center}
\end{figure}

The type of a knot is denoted by $C_n$ where the integer $C$ counts the minimum 
number of crossings and the subscript $n$ distinguishes knots that differ 
topologically \cite{kauffman1991}. For example, the sketch in 
Fig.\ \ref{fig:closure_scheme} shows the trefoil knot 3$_1$.
In a numerical analysis this classification
can be determined by computing topological invariants such as the Alexander 
polynomial $\Delta(t)$ \cite{kauffman1991}. More precisely we actually 
computed the knot parameter
\begin{eqnarray}
D = \Delta_p(-1.1)
\label{eq:D}
\end{eqnarray}
where $\Delta_p(t)=\lvert \Delta(t) \times \Delta(1/t) \rvert$
removes undesired prefactors and the choice of the argument
$t=-1.1$ is justified empirically \cite{virnau2005knots,virnau2010}.
It is well known that even though for complicated knots the Alexander polynomial 
and hence also $D$ is not unique [e.g., $D(5_1)=D(10_{132})$] \cite{kauffman1991}, 
all the simple knots relevant in this work can be distinguished by the 
parameter (\ref{eq:D}); see Table 1 of Ref.\ \cite{macromolecules21}. For
instance, $D=1$ for an unknotted chain and $D \approx 9.05463$, 25.09099, 25.45745, 
or 9.72667 for a chain with a 3$_1$, 4$_1$, 5$_1$, or 8$_{19}$ knot.
For a schematic illustration see Fig.\ \ref{fig:knots}.

\section{Results}
\label{sec:results}

In the analysis of our Monte Carlo data, while
mainly focusing on the behavior of the knot parameter, we have 
constructed in Ref.\ \cite{macromolecules21} the entire 
phase diagrams of the bead-spring and bead-stick models 
in the ($T,\kappa$) plane for several ratios 
$r_b/r_{\rm min}$. To this end we performed
elaborate simulations at many $T$ and $\kappa$ points for
chain length $N=14$ (and less extensively also for $N=28$)
and determined beside the 
average
knot parameter 
$\langle D \rangle$ 
also several other
observables such as 
the average squared radius of gyration $\langle R^2_g \rangle$, 
its temperature derivative, the average energies $\langle E_0 \rangle$, $\langle E_1 \rangle$, and
$\langle E \rangle = \langle E_0 \rangle + \kappa \langle E_1 \rangle$, 
as well as
the specific heat $C_v = \mathrm{d} \langle E \rangle/\mathrm{d} T
= (1/T^2) (\langle E^2 \rangle - \langle E \rangle^2)$. 
By means of the weighted histogram analysis method (WHAM), this 
information was finally combined to arrive at a quantitative
version of the generic phase diagram sketched in 
Fig.\ \ref{fig:generic_phase-diagram}.

\begin{figure}[h]
\begin{center}
\includegraphics[scale=.60]{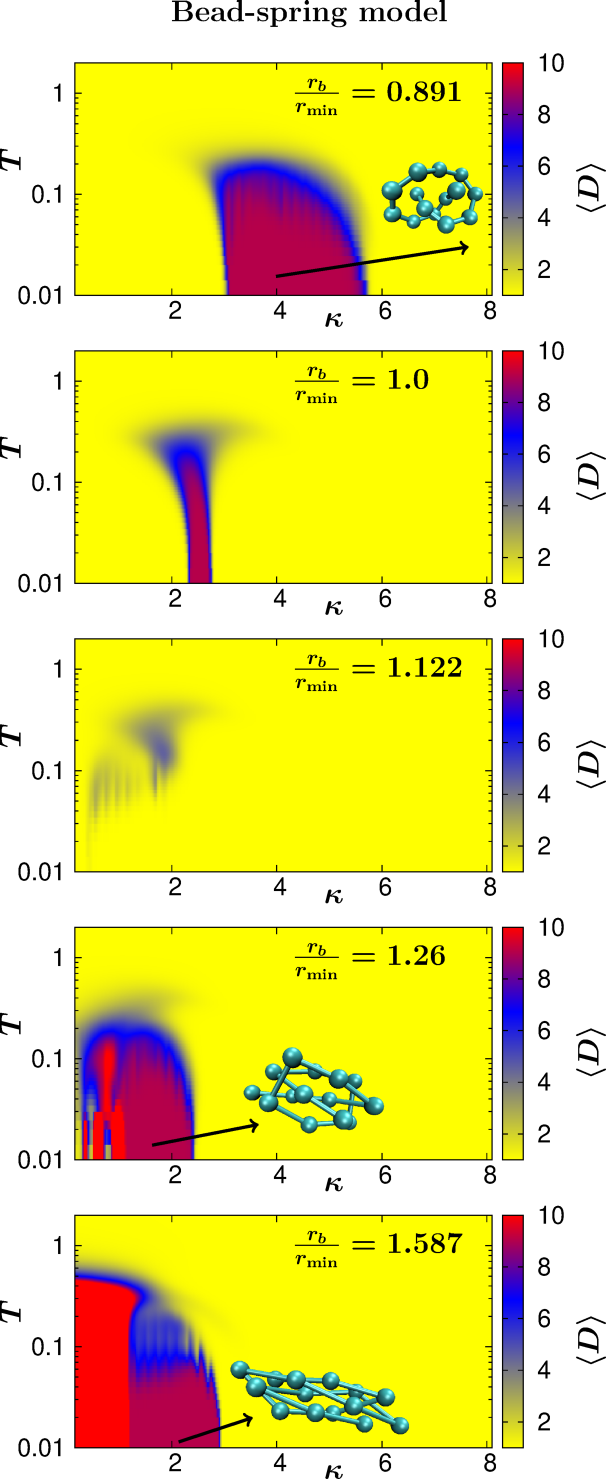}\hspace*{5mm}
\includegraphics[scale=.60]{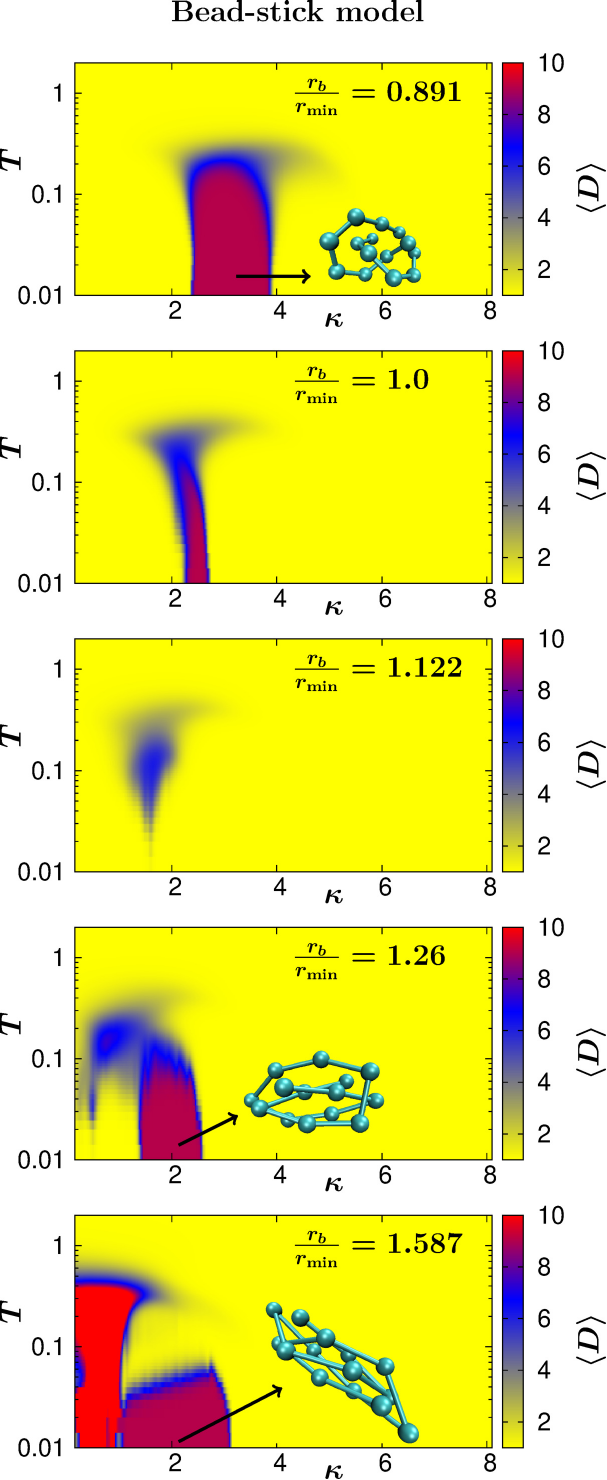}
%
%
\caption{
Phase diagram for a semiflexible polymer of length $N=14$ in the ($T,\kappa$)
plane with the average knot parameter $\langle D \rangle$ as order parameter and
several choices of the ratio $r_b/r_{\rm{min}}$ using a bead-spring (left) and
bead-stick (right) model. The snapshots show typical polymer conformations in 
the phase characterized by a stable trefoil knot $3_1$.
Taken from Ref.\ \cite{macromolecules21}.
}
\label{fig:knot-heat-map}       
\end{center}
\end{figure}

In Fig.\ \ref{fig:knot-heat-map} we focus only on the 
average
knot parameter
$\langle D \rangle$ 
which is shown in surface plots for the bead-spring (left) and
bead-stick (right) models with chain length $N=14$ for five different
values of the ratio\footnote{ Note that 0.891 is a lower bound since
otherwise the bond length $r_b$ would be smaller than the monomer
diameter $\sigma$.
}
$r_b/r_{\rm{min}}$ ($ =
2^{-1/6} \approx 0.891,
1.000,
2^{1/6} \approx 1.122,
2^{2/6} \approx 1.260$, and
$2^{4/6} \approx 1.587$).
The dark red color indicates regions in the phase diagram that
are populated by conformations forming a trefoil knot 3$_1$ with
$D \approx 9.05463$, whereas yellow color signals
unknotted (e.g., bent or extended) conformations. The blue boundary
layer results from phase coexistence of a finite system [$D \approx
(9.05+1)/2 \approx 5)$], for more details see below.

For the smallest ratio $r_b/r_{\rm{min}} = 2^{-1/6} \approx 0.891$
the knot phase is rather pronounced and seems to be a bit wider for
the bead-spring than the bead-stick model. In both models, it
significantly shrinks for $r_b/r_{\rm{min}} = 1$, and for
$r_b/r_{\rm{min}} = 2^{1/6} \approx 1.122$ it basically has disappeared.
By estimating the energies needed for forming a 3$_1$ knot and a bent
conformation (with three segments), respectively, one concludes that
bent conformations are more favorable in this case (the temperature
is so low that entropic contributions can be safely neglected)
\cite{macromolecules21}.

For larger ratios, the K3$_1$ knot phase reoccurs at smaller values
of $\kappa$. The main difference is that here the knot geometry is
flat while for $r_b <  r_{\rm min}$ it is spherical \cite{macromolecules21}.
The light red (or orange) region for $r_b/r_{\rm min} = 1.587$ at small $\kappa$
extending to relatively high $T$ signals a mixed phase of 4$_1$ knots and
unknotted conformations encoded by 
$D \approx (25.09+1)/2 \approx 13$
(it does not correspond to the 8$_{19}$ knot with 
$D \approx 9.72667$ 
since the $N=14$ chain is too short to accommodate a knot with 8 crossings).

As already observed in Ref.\ \cite{marenz2016knots} and mentioned above, 
most of the low-temperature
(pseudo) phase transitions exhibit a fairly strong first-order
signal which 
is caused by
the pronounced coexistence of
different conformational motifs (e.g., bent and knotted).
This
explains why the computer simulations are so demanding
and require the application of quite elaborate sophisticated methods
\cite{janke_rugged}. 
In the following 
we shall mainly focus on the 
blue colored region of
transitions between the 
bent (B) and knot (K) phase when $\kappa$ is varied at low temperature,
cf.\ Figs.\ \ref{fig:generic_phase-diagram} and \ref{fig:knot-heat-map}. 
For $N=14$ and 28, the specific bent phase is of type D3 in 
both cases, featuring 
bent conformations with 3 segments, and the knot phases are 
characterized by 3$_1$ and 5$_1$ knots, respectively. 
Note that for larger $N$, the phase 
diagram
is much richer and 
the picture becomes more complicated. For instance, with $N=42$ 
we observed transitions between the bent phase of type D6 and 
the K7$_1$ knot 
phase at low temperature respectively with the K8$_5$ knot phase at very low 
temperature \cite{marenz-wj-tobe}.
 
\begin{figure}[tb]
\begin{center}
\includegraphics[scale=.235]{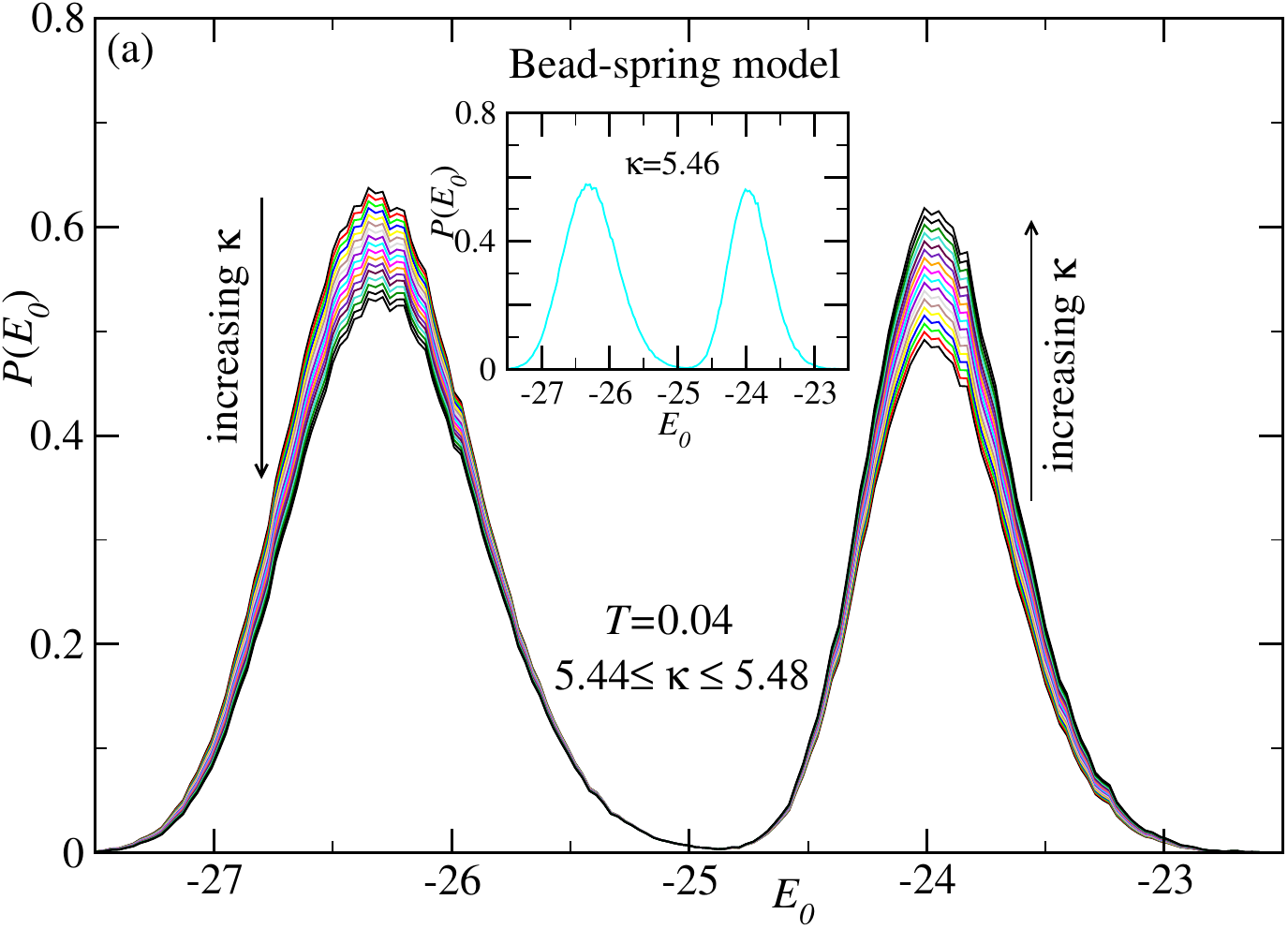}\hspace*{5mm}
\includegraphics[scale=.235]{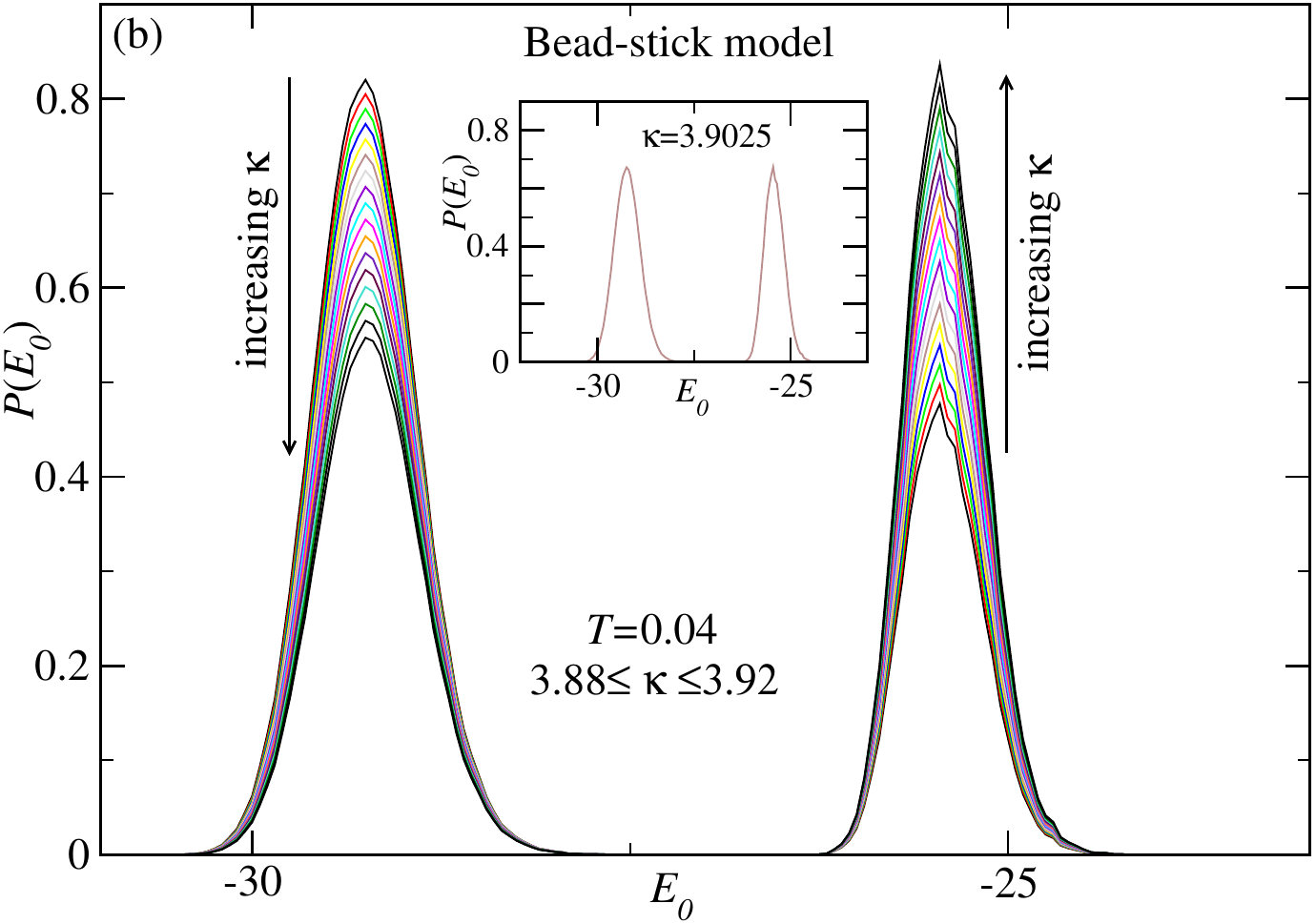}\vspace*{4mm}
\includegraphics[scale=.235]{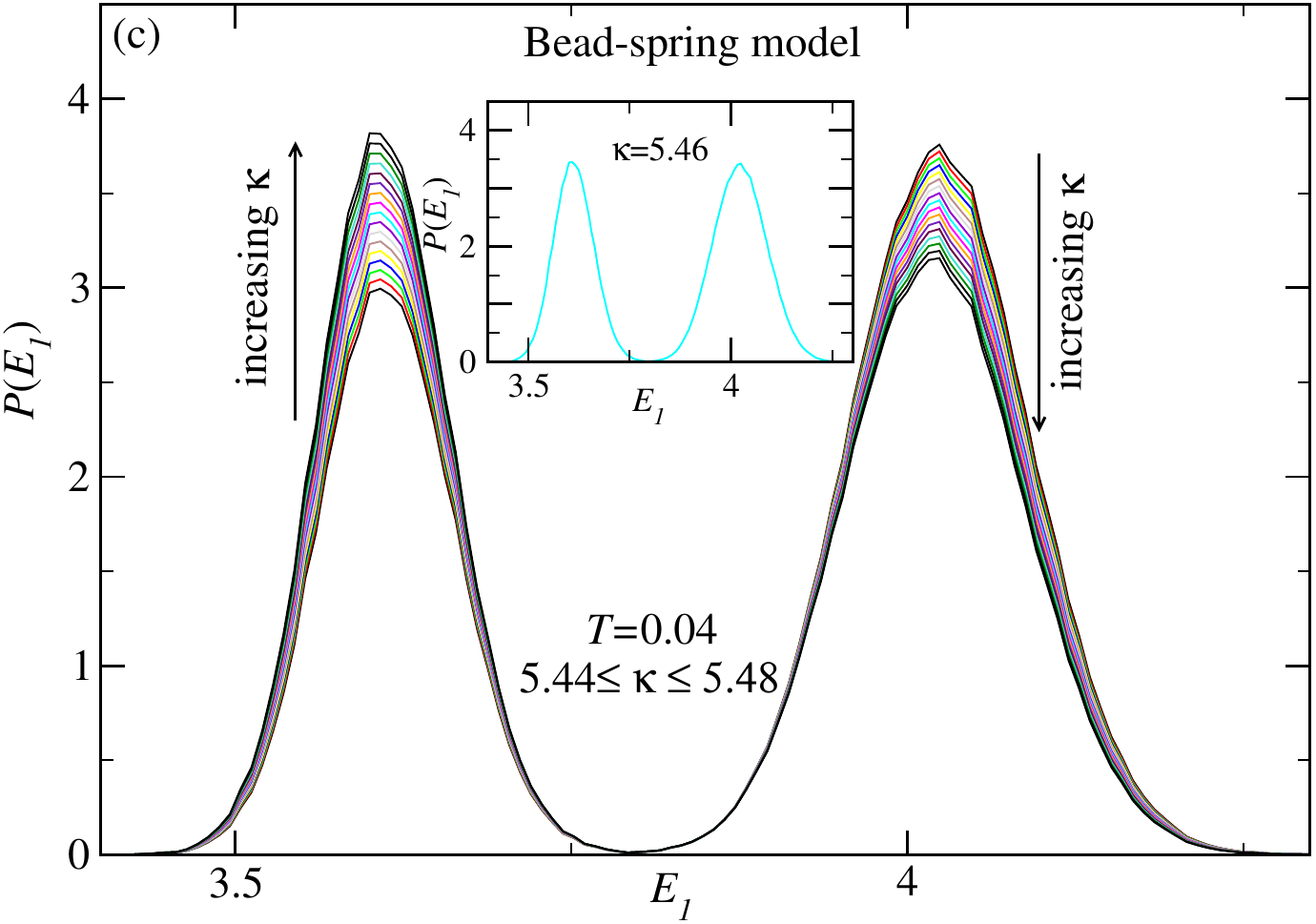}\hspace*{5mm}
\includegraphics[scale=.235]{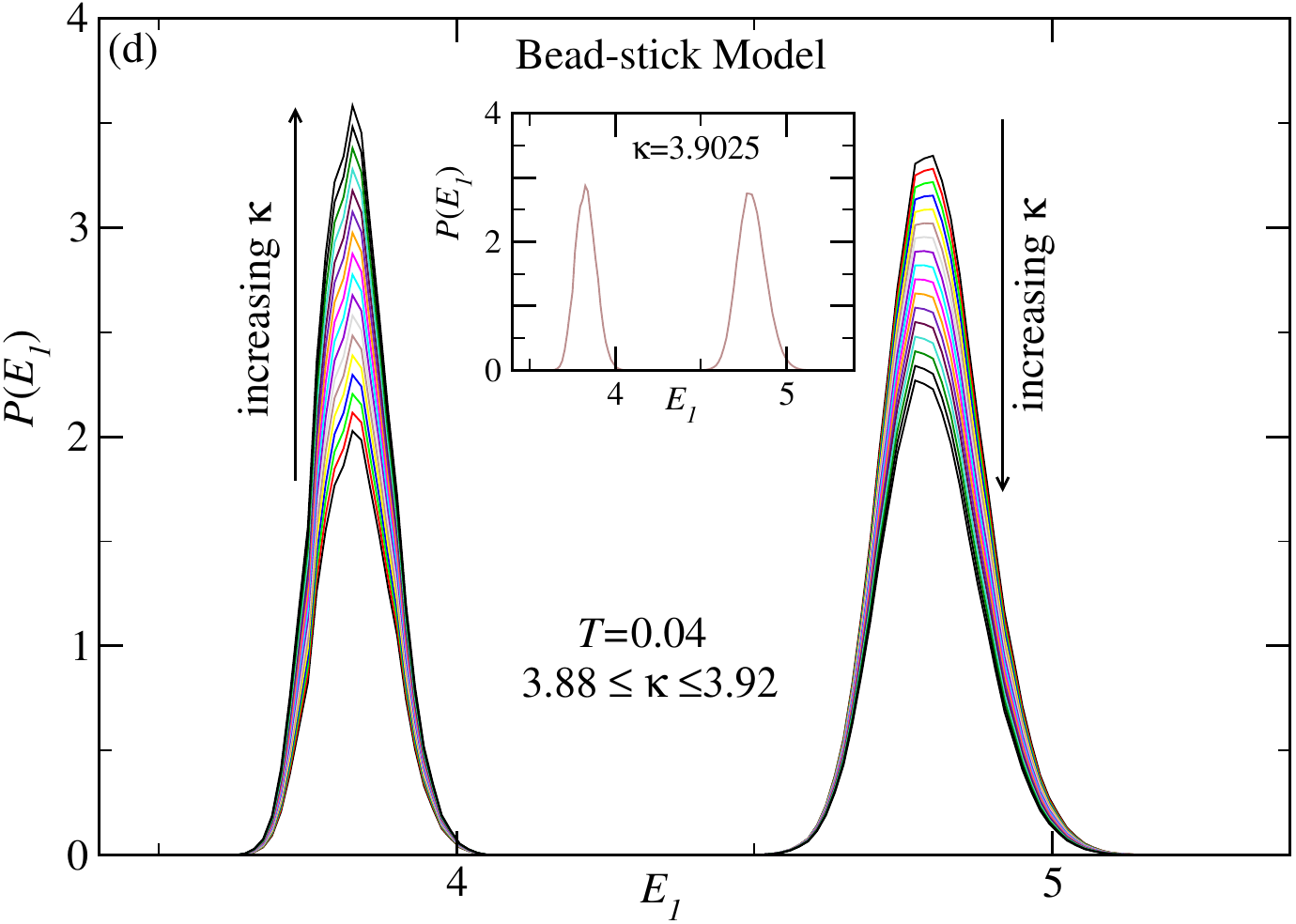}
\caption{Probability distributions $P(E_0)$ and $P(E_1)$ of the two 
contributions $E_0 = E_{\rm nb} + E_{\rm FENE}$ and 
$E_1 = E_{\rm bend}/\kappa$ to the total energy $E = E_0 + \kappa E_1$
in dependence of $\kappa$
close to the transition between the K3$_1$ knot phase and 
the bent phase of type D3 at $T=0.04$. The left panels (a), (c)
show our results 
for the bead-spring model and the right panels (b), (d) those for the
bead-stick model (where $E_{\rm FENE} \equiv 0$) with 
$r_b/r_{\rm min} = 0.891$ and chain length $N=14$. 
The insets concentrate on the phase-transition point $\kappa_t$,
defined from the distribution for which the two peaks have equal height.
Computing the ratio of maximum to minimum 
and taking the logarithm yields estimates of the free-energy barriers
at the transition (cf.\ Table \ref{tab:free-energy}).
}
\label{fig:free-energy}       
\end{center}
\end{figure}

These structural transitions turned out to be very intriguing since 
they do not carry any obvious first-order signal in the total energy 
$E = E_{0} + \kappa E_{1}$; 
rather, the distribution $P(E)$ looks apparently single-peaked 
which 
is
the typical signature of a smooth second-order 
transition. Only when looking at the two-dimensional distribution 
$P(E_{0},E_{1})$ a bimodal structure becomes clearly visible \cite{marenz2016knots}
which is typical for a first-order-like transition. The
double-peak 
structure is also shared by the individual one-dimensional 
distributions $P(E_0)$ and $P(E_1)$ separately. This is illustrated in 
Fig.\ \ref{fig:free-energy} for the $N=14$ chain with 
$r_b/r_{\rm min} = 0.891$ which shows for $T=0.04$ how the double peak varies 
as the bending stiffness $\kappa$ changes across the transition point $\kappa_t$.
In $P(E_0)$ the left peak corresponds to the knot phase and the 
right peak to the bent phase, whereas for $P(E_1)$ this is just
reversed.
The insets highlight the situation where both peaks are of equal 
height which is usually used as a 
criterion
for locating the
phase-transition point \cite{wj_prb93,janke_1st-order-review}.

The ratio of maximum to minimum of the equal-height probability 
distribution determines the strength or ``hardness'' of a first-order 
transition \cite{wj_prb93,janke_1st-order-review},
which is directly related to the corresponding free-energy barrier
per monomer
$\Delta f = T \ln \left(P_{\rm max}/P_{\rm min}\right)/N$
\cite{janke_2017,zierenberg_natcomm}.
The quantitative results for the two examples in Fig.\ \ref{fig:free-energy} 
are compiled in Table \ref{tab:free-energy}. The much larger ratios for the 
bead-stick model indicate that with fixed bonds the K3$_1$ knot phase is 
considerably more stable against forming bent conformations of D$3$ type
than with flexible springs. This is intuitively expected since the 
deformation of conformational motifs is easier with flexible springs 
than with stiff bonds.

The results presented here 
in Fig.\ \ref{fig:free-energy} and Table \ref{tab:free-energy}
are only of explorative nature.
We are currently investigating the shapes of the energy distributions 
and the associated free-energy barriers also at
several other structural transitions and in particular for
larger chain lengths $N$. This is a rather ambitious 
endeavor which requires fairly long simulation and analysis times, 
and hence further more extensive results will be reported 
elsewhere \cite{sm_sp_wj-tobe}.

\begin{table}[tb]
\begin{center}
\caption{Parameters of the double-peak distributions $P(E_0)$
and $P(E_1)$ of a semiflexible polymer with $N=14$ beads shown 
in the insets of Fig.\ \ref{fig:free-energy} at the transitions 
D3 $\leftrightarrow$ K3$_1$ for $T=0.04$ at $\kappa_t$. ``ratio''
stands short for the ratio $P_{\rm max}/P_{\rm min}$. 
}
\begin{tabular}{p{1.6cm}p{1.2cm}p{1.2cm}p{1.7cm}p{1.3cm}p{1.2cm}p{1.7cm}p{0.8cm}}
\hline\noalign{\smallskip}
model & $\kappa_t$ & $P_{\rm max}(E_0)$ & $P_{\rm min}(E_0)$ & ratio & $P_{\rm max}(E_1)$ & $P_{\rm min}(E_1)$ & ratio \\
\noalign{\smallskip}\svhline\noalign{\smallskip}
bead-spring & 5.46   & 0.58 & 0.003 & 193 & 3.61 & 0.01 & 361 \\
bead-stick  & 3.9025 & 0.67 & $4.26 \times 10^{-5}$ & 15\,728 & 2.88 & $4.02 \times 10^{-4}$ & 7\,164 \\
\noalign{\smallskip}\hline\noalign{\smallskip}
\end{tabular}
\label{tab:free-energy}
\end{center}
\end{table}

%
%
%
%
%
%
 

%

\section{Discussion}
\label{sec:discussion}

As we have seen in Figs.\ \ref{fig:knot-heat-map} and \ref{fig:free-energy},
semiflexible bead-spring and bead-stick models behave qualitatively very 
similarly and both exhibit the same generic phase diagram depicted 
in Fig.\ \ref{fig:generic_phase-diagram}. So, why did Seaton {\em et al.\/} 
\cite{seaton2013flexible} not identify knots in their seminal study of a 
bead-spring model? There are several related reasons:
\begin{enumerate}[label={\em \roman*\/})]
\item
In this first study of the entire phase diagram for 
a semiflexible polymer, they
focused on bent, toroidal, etc.\ motifs, but did not actively search for knots.
\item
The parameterization employed in Ref.\ \cite{seaton2013flexible}
realizes the ratio $r_b/r_{\rm min} = 1$ where only a 
very small 
region of the phase 
diagram is populated with knots. Moreover,
knots 
predominantly occur at very low temperatures below the lower limit of their study.
\item
The parameterization of the bead-spring model in 
Ref.\ \cite{seaton2013flexible} is a bit uncommon.
We have shown that it corresponds in good approximation to our 
bead-spring model, albeit with an extremely large spring constant of $K = 297.5$
\cite{macromolecules21}.
With additional simulations we have explicitly verified \cite{macromolecules21} 
that for such a large spring constant, as expected, the bead-spring model is 
basically indistinguishable from the bead-stick model.
\end{enumerate}

\vspace*{-1mm}
\section{Conclusions}
\label{sec:conclusions}

Our extensive two-dimensional replica-exchange Monte Carlo simulations 
of bead-stick and bead-spring homopolymer models confirm the theoretical
expectation that the existence of stable knotted phases depends on the 
ratio $r_b/r_{\rm min}$ between the equilibrium bond length $r_b$ and 
the distance $r_{\rm min}$ of the strongest attraction of nonbonded beads. 
By simple energy arguments one can read off that for $r_b/r_{\rm min} \approx 1$ 
the alternative bent motifs are more favorable than knotted structures.
This is compatible with our study of a semiflexible polymer adsorbed on a
flat surface where with the choice $r_b/r_{\rm min} = 1$ (and not very 
low temperatures $T \ge 0.1$) we did not observe knots \cite{austin2017interplay}.
This is probably also the main explanation why no knots have been reported
in the simulation study of Seaton {\em et al.} \cite{seaton2013flexible}.

The transitions between the various structural motifs at low
temperatures are first-order-like with pronounced phase coexistence 
at the transition point. In the energy distributions, this is reflected 
by a double-peak structure which we exploit for estimating the free-energy
barriers.
The results of the exploratory study described here are promising and
a more detailed investigation will be reported elsewhere \cite{sm_sp_wj-tobe}.

\vspace*{-1mm}
\begin{acknowledgement}
This work was funded by the Deutsche Forschungsgemeinschaft (DFG, German Research Foundation)
under Grant No.\ 189\,853\,844--SFB/TRR 102 (project B04) and further supported by the Deutsch-Franz\"osische 
Hoch\-schule (DFH-UFA) through the Doctoral College ``$\mathbb{L}^4$'' under Grant No.\ CDFA-02-07
and the Leipzig Graduate School of Natural Sciences ``BuildMoNa''. 
S.M.\ thanks the Science
and Engineering Research Board (SERB), Govt.\ of India for
a Ramanujan Fellowship (File No.\ RJF/2021/000044).
S.P. acknowledges support of the ICTS-TIFR, DAE, 
Govt.\ of India,
under Project No.\ RTI4001 
through
a research fellowship.

\end{acknowledgement}
%

%
%


\end{document}